\documentclass{elsart}
\usepackage{epsfig}
\usepackage{amssymb}
\begin{document}
\begin{frontmatter}
\title{A mechanism to synchronize fluctuations in scale free networks using growth models.}
\author[unmp]{C. E. La Rocca},
\author[unmp]{A. L. Pastore y Piontti},
\author[unmp,cps]{L. A. Braunstein},
\author[unmp]{P. A. Macri}.

\address[unmp]{Instituto de Investigaciones F\'isicas de Mar del Plata (IFIMAR)-Departamento de F\'isica, Facultad de Ciencias Exactas y Naturales, Universidad Nacional de Mar del Plata-CONICET, Funes 3350, (7600) Mar del Plata, Argentina.}
\address[cps]{Center for Polymer Studies, Boston University, Boston, Massachusetts 02215, USA}

\begin{abstract}

In this paper we study the steady state of the fluctuations of the surface
for a model of surface growth with relaxation to any of its lower nearest
neighbors (SRAM) [F. Family, J. Phys. A {\bf 19}, L441 (1986)] in scale free
networks. It is known that for Euclidean lattices this model belongs to the
same universality class as the model of surface relaxation to the minimum
(SRM). For the SRM model, it was found that for scale free networks with
broadness $\lambda$, the steady state of the fluctuations scales with the
system size $N$ as a constant for $\lambda \geq 3$ and has a logarithmic
divergence for $\lambda < 3$ [Pastore y Piontti {\it et al.}, Phys. Rev. E
{\bf 76}, 046117 (2007)]. It was also shown [La Rocca {\it et al.},
Phys. Rev. E {\bf 77}, 046120 (2008)] that this logarithmic divergence is due
to non-linear terms that arises from the topology of the network. In this
paper we show that the fluctuations for the SRAM model scale as in the SRM
model. We also derive analytically the evolution equation for this model for
any kind of complex graphs and find that, as in the SRM model, non-linear
terms appear due to the heterogeneity and the lack of symmetry of the
network. In spite of that, the two models have the same scaling, but the SRM
model is more efficient to synchronize systems.
\end{abstract}

\begin{keyword}
complex networks \sep interface growth models \sep transport in complex networks.

\PACS  89.75.-k \sep 89.20.-a \sep 82.20.Wt \sep 05.10.Gg
\end{keyword}
\end{frontmatter}

Recently, much effort has been devoted to the study of dynamics in complex
networks.  This is because many physical and dynamic processes use complex
networks as substrates to propagate, such as epidemic spreading
\cite{Pastorras_PRL_2001}, traffic flow
\cite{Lopez_transport,zhenhua,Guanliang} and synchronization
\cite{Jost-prl,Korniss07}. In particular, synchronization problems in
networks are very important in many fields such as the brain network
\cite{JWScanell}, networks of coupled populations in epidemic outbreaks
\cite{eubank_2004} and the dynamics and fluctuations of task completion
landscapes in causally-constrained queuing networks \cite{Kozma05}.
Synchronization deals with the optimization of the fluctuations in the steady
state of some scalar field $h$, that can represent the neuronal population
activity in brain networks, infected population in epidemics and jobs or
packets in queuing networks. It is particularity interesting to understand
how to reduce the load excess in communication networks in the steady
state. This problem can be mapped into a problem of non-equilibrium surface
growth where $h$ is a random scalar field on the nodes that could represent
the total flow on the network and the load excess could represent the
overload of flow that a node should handle. Then, a way to reduce the load
excess is to reduce the fluctuations of that scalar field. Recently, Pastore
y Piontti {\it et. al} \cite{anita} used the model of surface relaxation to
the minimum (SRM), that allow balancing the load, reducing the fluctuations
in scale free (SF) networks with degree distribution given by $ P(k) \sim
k^{-\lambda}$, ($k \ge k_{min}$) with $k$ the degree of a node, $k_{min}$ the
minimum degree that a node can have, and $\lambda$ the broadness of the
distribution \cite{Barabasi_sf}. Given a scalar field $h$ on the nodes, that
in surface problems represents the interface height at each node, the
fluctuations are characterized by the average roughness $W(t)$ of the
interface at time $t$, given by $W \equiv W(t) = \left\{1/N \sum_{i=1}^N
(h_i-\langle h \rangle)^2\right\}^{1/2},$ where $h_i \equiv h_i(t)$ is the
height of the node $i$ at time $t$, $ \langle h \rangle
=(1/N)\sum_{i=1}^Nh_i$, $N$ is the system size, and $\{ . \}$ denotes average
over configurations.

The aim of this paper is to study the steady state of $W \equiv W_s$ for the
model of surface growth with relaxation to any of its lower nearest
neighbors (SRAM) \cite{family}. We find that this model has the same behavior with the
system size as the SRM model for every $\lambda$, even though the SRM model
is more efficient to reduce the fluctuations and to enhance synchronization
than the SRAM model, as we show later. Moreover, we derive analytically the
general evolution equation for the SRAM model for any kind of random graph.

\section{Surface relaxation models}
In the SRM model, at each time step a node $i$ is chosen with
probability $1/N$. If we denote by $v_i$ the nearest-neighbor
nodes of $i$, then

\[ \mbox{if}  \left  \{\begin{array}{ll}
  (1)\ h_i \leq h_j,\ \forall j \in v_i & \Rightarrow h_i=h_i+1 \\
  (2)\ h_j < h_n\ \forall n \not= j \in v_i & \Rightarrow
h_j = h_j+1.
\end{array}\right.
\]

The SRAM model has the same rule (1) as the SRM model, but the second rule is
different: the chosen node can relax to any of its lower $m$ neighbors with
probability $1/m$. Then, the rules for the SRAM model are
\begin{eqnarray}\label{Regpep}
  \mbox{if}\ \left  \{\begin{array}{ll}
  (1)\ h_i \leq h_j,\ \forall j \in v_i & \Rightarrow h_i=h_i+1, \mbox{else} \\
  (2)\ \exists\ m\ \mbox{nodes}\ j \in v_i\ \mbox{with}\ h_j < h_i & \Rightarrow
h_j=h_j+1\ \mbox{with probability}\ 1/m.
\end{array}\right.
\end{eqnarray}

It is known that in Euclidean lattices of typical linear size $L$ the SRM and
the SRAM models belong to the same universality class \cite{family,Vveden}, with

\[ W(t) \sim \cases {
    \begin{array}{rl} t^\beta, & t < t_s \ ,\\
        L^\alpha, & t > t_s \ ,\end{array} }
\]
where $t_s$ is the saturation time which scales as $t_s \sim L^z$. Here, the exponent
$\beta$ is the growth exponent, $\alpha$ is the roughness exponent and $z$ is
the dynamical exponent that characterizes the growth correlations given by
$z=\alpha/\beta$. For $1$ dimension, these exponents are $\beta=1/4$,
$\alpha=1/2$ and $z=2$. Moreover, both growth models belong to the same universality
class as the Edward-Wilkinson (EW) equation,

\[
  \frac{\partial h(x,t)}{\partial t}= \nu \nabla^2 h(x,t) + \eta(x,t)\ ,
\]
where $\nu$ is the coefficient of surface tension and
$\eta(x)$ is a Gaussian noise with zero mean and
covariance given by

\[
\{\eta(x,t) \eta(x^{'},t^{'})\} = 2 D
\delta(x-x^{'})\delta(t-t^{'})\ .
\]
Here, $D$ is the diffusion coefficient and is taken as a constant. The fact
that these models are represented by the same phenomenological equation is
due to the symmetry of those models on the underlying Euclidean substrate
\cite{barabasi-stanley}. The extension of the EW equation to any unweighted
graph with $N$ nodes is described by
\begin{equation}\label{Eq.ew}
  \frac{\partial h_i(t)}{\partial t}= \nu \sum_{j=1}^N A_{ij} (h_j(t) -h_i(t) ) + \eta_i(t)\ ,
\end{equation}
where $i$ and $j$ are nodes of the graph, $\{ A_{ij}\}$ is the
adjacency matrix ($A_{ij}=1$ if $i$ and $j$ are connected and zero
otherwise), $\nu$ represents the same as in Euclidean lattices and
$ \eta_i (t)$ is a white Gaussian noise with zero mean and covariance
given by

\[ \{\eta_i(t) \eta_j(t^{'})\} = 2 D\ \delta_{ij}\delta(t-t^{'})\ ,\]
being $D$ the same as in Euclidean lattices.

In Ref.~\cite{anita} it was found that the saturation regime of $W_s$ in SF
networks scales with $N$ as
\begin{equation} W_s \sim \cases {
    \begin{array}{ll} const. & \mbox{for}\ \lambda \ge 3 , \\
        \ln N & \mbox{for}\ 2 <\lambda < 3\ . \end{array} }.
\end{equation}

It was also shown that the EW equation given by Eq. (\ref{Eq.ew}) predicts
that in the thermodynamic limit $W_s \sim 1/\langle k \rangle$ for any random
graph \cite{Korniss07}. Then, the unweighted EW equation in random graphs
cannot describe the SRM model. La Rocca {\it et. al} \cite{cristian}, using a
temporal continuous approach, derived the evolution equation that describes
the SRM model. They found that the logarithmic divergence for $\lambda < 3$
cannot be explained by the unweighted EW equation [See Eq. (\ref{Eq.ew})] in
graphs. The equation derived in Ref \cite{cristian} contains non-linear terms
and weights that appear as a consequence of the heterogeneous topology that a
SF has for $\lambda <3$, even though the network is unweighted. The
heterogeneity breaks the symmetry $h \to -h$ of Eq.~(\ref{Eq.ew}). For
$\lambda \geq 3$ the heterogeneity is not strong enough and the non-linear
terms are negligible for the system sizes studied there, and the behavior of
the fluctuations becomes well described by a weighted EW equation. It is not
unexpected that transport processes in random heterogeneous graphs behave
differently than in Euclidean lattices due to the fact that the nodes with
high degrees (hubs) play a mayor role in transport. For example,
reaction-diffusion processes behave very differently in homogeneous lattices
than in SF networks with $2<\lambda<3$ due to the presence of hubs which
control the behavior for long times \cite{panos,boguna}. The hubs are
responsible of a superdiffusive regime because they diminish the distances.

\section{Saturation results for the SRAM model}

We construct our networks using the Molloy-Reed (MR) algorithm \cite{Molloy},
with $k_{min}=3$ in order to ensure that the network is fully connected. The
initial conditions for the scalar field $\{ h \}$ were drawn from a random
uniform distribution between $[0,1]$. In Fig.~\ref{fig.1}, we plot $W^2$ as a
function of $t$ for $\lambda=3.5$ and $2.5$ and different values of $N$. In
the insets of Fig.~\ref{fig.1} we plot $W_s$ as a function of $N$. We can see that for
$\lambda=3.5$, $W_s$ increases, but asymptotically goes to a constant and all
the $N$ dependence is due to finite-size effects as in the SRM model
\cite{anita}. However, for $\lambda=2.5$ we find that $W_s \sim \ln N$, as
in the SRM model. This is shown in the inset of Fig.~\ref{fig.1}$(b)$ in
log-linear scale.  Then, $W_s$ for both models scales in the same way for SF
networks, with a logarithmic divergence for $\lambda<3$ and as a constant for
$\lambda \geq 3$.

\section{Analytical Evolution Equation}

Next, we derive analytically the evolution equation for $\{ h \}$ of the SRAM
model for any kind of random graphs.

The procedure chosen here is the same as the one used in Ref.~\cite{cristian}
and is based on a coarse-grained (CG) version of the discrete Langevin
equations obtained from a Kramers-Moyal expansion of the master equation
\cite{VK,Vveden,lidia}. The discrete Langevin equation for the evolution of
the height in any growth model is given by \cite{Vveden,lidia}
\begin{eqnarray}\label{eqh}
\frac{\partial h_i}{\partial t}= \frac{1}{\tau}G_i + \eta_i\ ,
\end{eqnarray}
where $G_i$ represents the deterministic growth rules that cause the evolution of
the node $i$, $\tau=N \delta t$ is the mean time to grow a layer of the
interface, and $\eta_i$ is a Gaussian noise with zero mean and covariance given
by \cite{Vveden,lidia}
\begin{equation}\label{ruido}
\{\eta_i(t)\eta_j(t')\}=\frac{1}{\tau}G_i\delta_{ij}\delta(t-t')\ .
\end{equation}

If $k_\gamma$ represents the degree of node $\gamma$, we can write $G_i$ more
explicitly as
\begin{equation}\label{eqreglas}
G_i = \omega_i + \sum_{j=1}^{N} A_{i j}\ (\omega_j^{1}+\omega_j^{2}+...+\omega_j^{k_j})\ ,
\end{equation}
where $\omega_i$ is the growth contribution by deposition on the node $i$
and $\omega_j^{m}$ is the growth contribution to the node $i$ by relaxation
from its neighbor $j$ with probability $1/m$, being $m$ the number of
neighbors of the node $j$ with smaller heights than the node $j$. Then,

\[ \omega_i= \prod_{j \in v_i} \left[ 1- \Theta(h_i-h_j) \right],\]

\[ \omega_j^{1}= \left[ 1- \Theta(h_i-h_j) \right] \prod_{n \in v_j,n\not=i}
            \left[ 1- \Theta(h_j-h_n) \right]\ ,\]

\[ \omega_j^{2}= \left[ 1- \Theta(h_i-h_j) \right] \prod_{n \in v_j,n\not=i,m}
            \left[ 1- \Theta(h_j-h_n) \right]\frac{1}{2}\left[ 1- \Theta(h_m-h_j) \right]\ ,\]

\[ \omega_j^{3}= \left[ 1- \Theta(h_i-h_j) \right] \prod_{n \in v_j,n\not=i,m,\ell}
            \left[ 1- \Theta(h_j-h_n) \right]\frac{1}{3}\left[ 1- \Theta(h_m-h_j) \right]\left[ 1- \Theta(h_\ell-h_j) \right]\ ,\]
\begin{center}
\[. \]
\[. \]
\[. \]
\end{center}
\[ \omega_j^{k_j}= \left[ 1- \Theta(h_i-h_j) \right] \prod_{n \in v_j,n\not=i}
            \left[ 1- \Theta(h_n-h_j) \right]\frac{1}{k_j}\ .\]
Here, $\Theta$ is the Heaviside function given by $\Theta(x)=1$ if $x \geq 0$
and zero otherwise, with $x=h_t-h_s \equiv \Delta h$. Without loss of
generality, we take $\tau=1$ and assume that the initial configuration of
$\{h_i\}$ is random. A schematic plot of the growing rules are shown in
Fig.~\ref{fig.2} where the rule (1) represent $\omega_i$, (2) represent
$\omega_j^1$ and (3) represent $\omega_j^2$.

In the CG version $x \rightarrow 0$; thus after expanding an analytical
representation of $\Theta(x)$ in Taylor series around $x=0$ to second order in
$x$, we obtain
\begin{eqnarray}\label{nonliner}
G_{i}&&=\ a^{k_i}\ +\ \sum_{j=1}^{N}C_{ij}\ +\ c_1a^{k_i-1}\ \sum_{j=1}^{N}A_{ij}(h_j-h_i)\nonumber\\
  &&+\ \frac{c_1}{a}\ \sum_{j=1}^{N}C_{ij}(h_j-h_i)\
  +\ \frac{c_1}{a}\ \sum_{j=1}^{N}T_{ij}\sum_{n=1,n\not=i}^{N}A_{jn}(h_n-h_j)\nonumber\\
  &&-\ \frac{c_2}{a}\ \sum_{j=1}^{N}C_{ij}(h_j-h_i)^2\ -\ a^{k_i-1}\left[c_2+\frac{c_1^2}{2a}\right]\sum_{j=1}^NA_{ij}(h_j-h_i)^2\nonumber\\
  &&+\ \sum_{j=1}^{N}\left[\frac{c_1^2}{2a^2}Q_{ij}-\frac{c_2}{a}C_{ij}\right]\left[\sum_{n=1,n\not=i}^{N}A_{jn}(h_n-h_j)^2\right]\nonumber\\
  &&+\ \frac{a^{k_i-2}c_1^2}{2}\left[\sum_{j=1}^NA_{ij}(h_j-h_i)\right]^2\
  +\ \frac{c_1^2}{a^2}\sum_{j=1}^{N}T_{ij}(h_j-h_i)\left[\sum_{n=1,n\not=i}^NA_{jn}(h_n-h_j)\right]\nonumber\\
  &&-\ \frac{c_1^2}{2a^2}\sum_{j=1}^{N}Q_{ij}\left[\sum_{n=1,n\not=i}^NA_{jn}(h_n-h_j)\right]^2\ ,
\end{eqnarray}
where $a=(1-c_0)$, $c_0$, $c_1$ and $c_2$ are the first three coefficients of the
expansion of the $\Theta(x)$ and 
\begin{eqnarray}\label{pesos1}
C_{ij}=&A_{ij}\left[\frac{2^{k_j}-1}{k_j}\right]a^{k_j}\ ,\nonumber\\
T_{ij}=&A_{ij}\left[\frac{2^{k_j}-k_j-1}{k_j(k_j-1)}\right]a^{k_j}\ ,\nonumber\\
Q_{ij}=&A_{ij}\left[\frac{2(1-2^{k_j})+k_j^2+k_j}{(k_j-2)(k_j-1)k_j}\right]a^{k_j}\ ,\nonumber
\end{eqnarray}
are different ``weights'' on the link $ij$ introduced by the dynamics.

In our equation the non-linear terms in the difference of heights arise as a
consequence of the lack of a geometrical direction and the heterogeneity of
the underlying network.  This result is very similar to the one found for
the SRM model in SF unweighted networks \cite{cristian}, where the non-linear
terms appear due to the heterogeneity of the network. 

For the noise correlation [See Eq.~(\ref{ruido})], up to zero
order in $\Delta h$ \cite{Vveden,lidia} we obtain $\{\eta_i(t)
\eta_j (t^{'})\}= 2 D(k_i)\delta_{ij} \delta(t-t')$ with
\begin{eqnarray}\label{coefdifus}
D(k_i)=\frac{1}{2}[\ a^{k_i} + \sum_{j=1}^{N}C_{ij}\ ].
\end{eqnarray}
Notice that all the coefficients of the equation depend on the connectivity
of node $i$, {\it i.e.}, on the topology of the underlying network. This
dependence on the topology is expressed as weights on the links of the
unweighted underlying network that appears only due to the dynamics on the
heterogeneous network. 

\section{Numerical results of the analytical evolution equation}

We numerically integrate our evolution equation for SF networks taking into
account the linear terms and the first non-linear term of
Eq.~(\ref{nonliner}). This is because when non-linear terms are considered,
the numerical integration algorithms we use has numerical
instabilities. This is still an open problem to be solved in the future. For
all our integrations we used the Euler method with the representation of the
Heaviside function given by $\Theta(x)=\{1+ \tanh[U(x+z)]\}/2$, where $U$ is
the width and $z= 1/2$ \cite{lidia}, and random initial conditions. With our
choice of the representation of the Heaviside function, we obtain:
$c_0=[1+\tanh(U/2)]/2$, $c_1=[1-\tanh^2(U/2)]\ U/2$, and
$c_2=[-\tanh(U/2)+\tanh^3(U/2)]\ U^2/2$. 

In Fig.~\ref{fig.3}, we plot $W^2$ as a function of $t$, obtained from the
integration of Eq.~(\ref{eqh}) with Eq.~(\ref{nonliner}) and $D(k_i)$ given
by Eq.~(\ref{coefdifus}) for $\lambda=3.5$ and $2.5$ and different values of
$N$, with $k_{min}=3$. For the time step integration we chose $\Delta t \ll
1/k_{max}$ according to Ref.~\cite{pasointeg}, where $k_{max} \sim
N^{1/(\lambda -1)}$ is the degree cutoff for the MR construction. In the
inset figures we plot $W_s$ as a function of $N$. We can see that for
$\lambda=3.5$, $W_s$ increases, but asymptotically goes to a constant and all
the $N$ dependence is due to finite-size effects. However, for $\lambda=2.5$
we found a logarithmic divergence of $W_s$ with $N$, as shown in the inset of
Fig.~\ref{fig.3}$(b)$ in log-linear scale. The fit of $W_s$ with a
logarithmic function for $\lambda=2.5$ shows the agreement between our
results and those obtained for the SRAM model in SF networks for $\lambda
<3$. Then, our equation reproduces correctly the behavior of $W_s$ for the
model for any $\lambda >2$. Notice that for $\lambda \geq 3$ and the system
sizes studied here, the non-linear terms do not contribute and the process can
be described by a weighted EW equation.

\section{Discussions}

The behavior of the SRAM and SRM models in the steady state are the same and
both evolution equations are in agreement with the fact that all the
coefficients depend on the connectivity of a node, {\it i.e.}, on the
topology of the underlying network, and the weights appear only due to the
dynamics on the heterogeneous network. Another similarity between both models
is that for $\lambda \geq 3$ the non-linear terms do not play any role for
the systems size studied here, then both process are well described by a
weighted EW equation. However, the equations for both models are different
\cite{cristian} and the main difference appears in the weights that produce
different values of $W_s$ among them. For the synchronization problem, the
SRM model is more efficient than the SRAM model, as can be understood from a
lower $W_s$ shown in Fig.~\ref{fig.4} where we plot $W^2$ as a function of
$t$ in log-log scale for the SRM and SRAM models for $N=1024$ for $(a)\
\lambda=3.5$ and $(b)\ \lambda=2.5$. We can see that the SRM model reaches
the saturation regime faster than the SRAM model and $W_s$ for the first one
is much lower than $W_s$ of the latter. This means that for the same system
size, the process that will be better for synchronizing is the SRM, since it
has less fluctuations on its scalar fields. These observations can be
explained as follow: in both models, the nodes with low degree control the
process all the time because they are more abundant and make a major
contribution to the growth of the hubs. Then, we expect that the growing
contribution of the hubs will be by relaxation from their neighbors with
lower degree because our networks are disassortative for $\lambda < 3$ (due
to the MR construction) \cite{bog_pastor}. This is because in the SRM model
at the initial stages the hubs grow faster than in the SRAM model. As the
hubs are more important in the SRM model than in the SRAM model, the
height-height correlation length should growth faster allowing to reach the
saturation regime earlier. Notice that in the SRM model the nodes relax always
to the minimum, while in the SRAM model the relaxation takes place at any
randomly chosen neighbor with smaller height (not necessarily the minimum) than
the chosen node. Then, if we have to chose one of these models as a
synchronization process, it is more efficient to use the SRM model.

\section{Conclusions}

In summary, we studied the SRAM model in SF unweighted networks and found
that for $\lambda \geq 3$, $W_s$ scales as a constant and the $N$ dependence
is due to finite-size effects, while for $\lambda < 3$ there is a logarithmic
divergence with $N$, the same as in the SRM model. Then, the SRAM and SRM
models still scale in the same way for SF networks. We derived analytically
the evolution equation for the SRAM model for any network and find that even
when the underlying network is unweighted, the dynamic introduces weights on
the links that depend on the topology of the network. This equation contains
non-linear terms and considering the linear and only the first non-linear
term in the integration of the evolution equation, we recover the scaling of
$W_s$ with $N$ for any $\lambda > 2$. And last but not least, we found that
even though the two models have the same scaling, for synchronization
problems the SRM model is more efficient because it reaches the steady state
faster than the SRAM model and its fluctuations are much lower.

\subsubsection*{Acknowledgments}

This work has been supported by UNMdP and FONCyT (Pict 2005/32353).

\begin{figure}
\includegraphics[width=0.55\textwidth]{fig1a.eps}
\vspace{0.2cm}
\includegraphics[width=0.55\textwidth]{fig1b.eps}
\caption{$W^2$ as a function of $t$ for the SRAM model: $(a)\ \lambda=3.5$
for $N=64$ ($\bigcirc$), $128$ ($\Box$), $256$ ($\diamond$), $512$
($\bigtriangleup$), $1024$ ($\bigtriangledown$), $1536$ ($\star$) and $2048$
($X$). $(b)\ \lambda=2.5$ for $N=1024$ ($\bigcirc$), $1280$ ($\Box$), $1536$
($\diamond$), $1792$ ($\bigtriangleup$), and $2048$ ($\bigtriangledown$). In
the inset figures we plot $W_s$ as a function of $N$ in symbols. The inset
figure of $(b)$ is in log-linear scale and the dashed line represents the
logarithmic fitting of $W_s$ with $N$.\label{fig.1}}
\end{figure}

\begin{figure}
\includegraphics[width=0.5\textwidth]{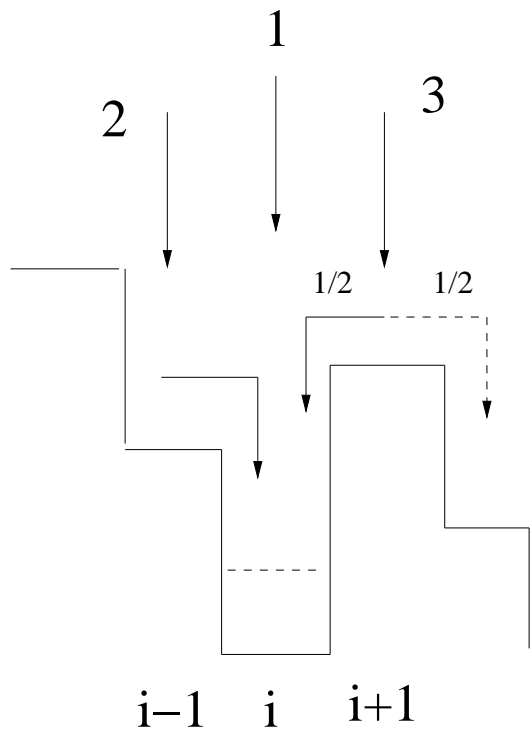}
\caption{Schematic plot of the growing rules for the SRAM model in
one-dimensional Euclidean lattice. The solid arrows indicate the
contributions to the growth of site $i$ due to deposition (1) and diffusion
from the nearest neighbors (2) and (3). Notice that in the case (3) site $i$
growths with probability $1/2$.
\label{fig.2}}
\end{figure}

\begin{figure}
\includegraphics[width=0.55\textwidth]{fig3a.eps}
\hspace{0.2cm}
\includegraphics[width=0.55\textwidth]{fig3b.eps}
\caption{$W^2$ as a function of $t$ for the integration of the evolution
equation using the linear terms and the first non-linear term in Eq.~(\ref{nonliner}): $(a)\
\lambda=3.5$ for $N=64$ ($\bigcirc$), $128$ ($\Box$), $256$ ($\diamond$),
$512$ ($\bigtriangleup$), $1024$ ($\bigtriangledown$) and $1536$
($\star$). $(b)\ \lambda=2.5$ for $N=192$ ($\bigcirc$), $256$ ($\Box$), $384$
($\diamond$), $512$ ($\bigtriangleup$), and $768$ ($\bigtriangledown$). In
the inset figure we plot $W_s$ as a function of $N$ in symbols. The inset figure of $(b)$
is in log-linear scale and the dashed line represents the fitting with
$W_s \sim \ln N$.\label{fig.3}}
\end{figure}

\begin{figure}
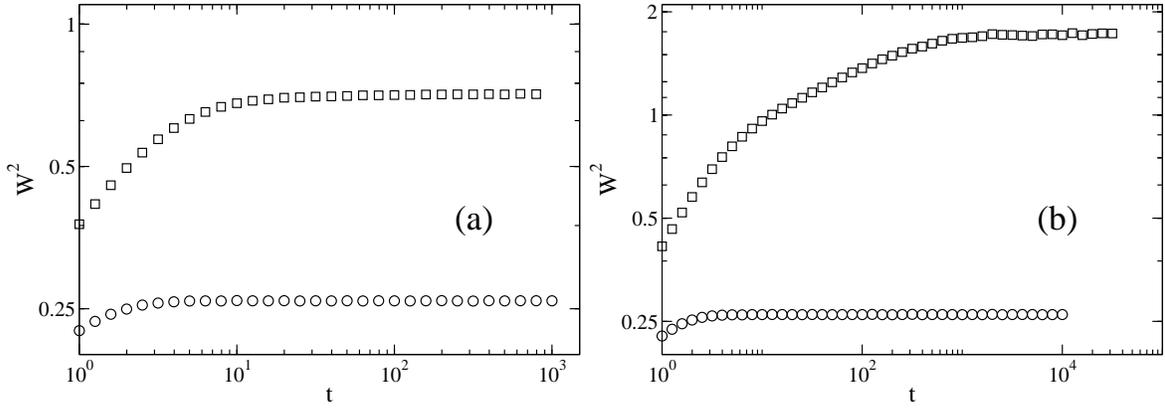

\includegraphics[width=0.55\textwidth]{fig4a.eps}
\vspace{0.2cm}
\includegraphics[width=0.55\textwidth]{fig4b.eps}
\caption{$W^2$ as a function of $t$ in log-log scale for the SRM ($\bigcirc$) and SRAM  ($\Box$)
models for $N=1024$ for: $(a)\ \lambda=3.5$ and $(b)\ \lambda=2.5$.\label{fig.4}}
\end{figure}

\end{document}